\documentclass[1p]{elsarticle}

\usepackage{lineno}
\usepackage{setspace}
\usepackage{xcolor}
\usepackage{graphicx}
\usepackage[labelfont=bf,labelsep=period,skip = 0pt]{caption}
\usepackage{xr-hyper}
\usepackage{hyperref}
\journal{Chemical Engineering Journal}

\begin{document}
\begin{doublespace}
\begin{frontmatter}

\title{Self-propelling Microdroplets Generated and Sustained by
	Liquid-liquid Phase Separation in Confined
	Spaces}

\author[1,2]{Xuehua Zhang\corref{cor1}}
\ead{xuehua.zhang@ualberta.ca}
\author[1,2]{Jae Bem You\fnref{fn1}}
\author[1]{Gilmar F. Arends\fnref{fn1}}
\author[1]{Jiasheng Qian\fnref{fn1}}
\author[2]{Yibo Chen}
\author[2,3]{Detlef Lohse}
\author[1]{John M. Shaw\corref{cor2}}
\ead{jmshaw@ualberta.ca}
\address[1]{Department of Chemical and Materials Engineering, University of Alberta, Alberta T6G 1H9, Canada}
\address[2]{Physics of Fluids Group, Max Planck Center Twente for Complex Fluid Dynamics, JM Burgers Center for Fluid Dynamics, Mesa+, Department of Science and Technology, University of Twente, Enschede 7522 NB, The Netherlands}
\address[3]{Max Planck Institute for Dynamics and Self-Organization, 37077 G\"{o}ttingen, Germany}

\fntext[fn1]{These authors contributed equally to this work.}
\cortext[cor1]{Corresponding author}
\cortext[cor2]{Corresponding author}

\begin{abstract}
Flow transport in confined spaces is ubiquitous in technological processes, ranging from separation and purification of pharmaceutical ingredients by microporous membranes and drug delivery in biomedical treatment to chemical and biomass conversion in catalyst-packed reactors and carbon dioxide sequestration. In this work, we suggest a distinct pathway for enhanced liquid transport in a confined space via self-propelling microdroplets. These microdroplets can form spontaneously from localized liquid-liquid phase separation as a ternary mixture is diluted by a diffusing poor solvent. High speed images reveal how the microdroplets grow, break up and propel rapidly along the solid surface, with a maximal velocity up to $\sim$160 $\mu m/s$, in response to a sharp concentration gradient resulting from phase separation. The microdroplet self-propulsion induces a replenishing flow between the walls of the confined space towards the location of phase separation, which in turn drives the mixture out of equilibrium and leads to a repeating cascade of events. Our findings on the complex and rich phenomena of self-propelling droplets suggest an effective approach to enhanced flow motion of multicomponent liquid mixtures within confined spaces for time effective separation and smart transport processes. 
\end{abstract}


\end{frontmatter}


\section{Introduction}

Microdroplets out of composition equilibrium can be seen as active matter, exhibiting autonomous motion without additional energy input from external sources \cite{lohse2020}. These self-propelling droplets have drawn intensive research interest, as they show promise as programmable microcarriers in drug delivery \cite{fluery2018}, artificial units simulating chemotaxis and collective behaviours in living microorganisms \cite{dreyfus2005,maass2016,Maass2018}, fast solvers of geometric mazes \cite{lagzi2010,jin2017pnas}, microswimmers to extract metal ions \cite{ban2014,ban2015}, or automated bouncing droplets for stirring of the surrounding liquid \cite{li2019prl-yanshen}, to name a few. The primary force driving autonomous microdroplet  motion is typically a Marangoni stress arising from imbalanced forces along droplet surfaces \cite{lohse2020}. The mobility of the droplets is governed by the interplay of droplet size, the viscosities of liquids inside and outside of droplet, and the interfacial tension gradient. 

The response of microdroplets to small gradients in interfacial tension leads to autonomous motion via concentration gradients of chemical constituents adjacent to the droplet surface. Strategies have been developed to sustain one component out of equilibrium so that the droplets can travel in a designated direction over a large distance for an extended period of time \cite{wang2019,banerjee2019,gao2014catalytic}. A typical approach is the creation of a non-uniform distribution of molecular components on a droplet surface, for instance, from dissolution of a solid surfactant source \cite{banerjee2019}, from local chemical and catalytic reactions \cite{boulogne2017,kar2014,cejkova2014} or solvent vapor emitted by neighbouring droplets \cite{cira2015,majhy2020}.  Even simple combinations of different types of oil and aqueous solutions are able to yield diverse behaviors of the droplets that can only be completely probed through robotic experimental platforms and machine learning \cite{points2018}. Droplet motion is further enhanced by coupling it with induced external liquid flow. As exemplified by a bouncing droplet in a stratified liquid\cite{li2019prl-yanshen}, Marangoni stress at the droplet surface pulls down the bulk liquid, further strengthening the stress. The droplet jumps up once the stress exceeds gravity. The process repeats itself as gravity restores the system \cite{li2019prl-yanshen}. 

So far creation of self-propelling motion has mainly focused on droplets in a continuous liquid medium or on liquid surfaces with small resistance to droplet movement. However, in the biological context or for technological processes, microdroplets evolve in compartmentalized environments.  Examples include cases as diverse as coacervate droplets of biomolecules causing liquid-liquid phase separation in organelles within living cells,\cite{hyman2014} and emulsion droplets in flooding fluid in enhanced oil recovery.\cite{zhou2019oil} In addition to significant extra drag resistance along the boundary acting on the droplets, the confined space inherently introduces effects from the size, composition and motion of droplets on their surroundings. To explore the potential for self-propelled droplets in confinement, it is  essential to identify an approach to sustain autonomous motion of droplets and to understand the coupled interactions between the droplets and their confined environment.

In this work, we reveal an effective pathway for rapid self-propelled transport of microdroplets in quasi-2D confined spaces. These microdroplets are secondary droplets formed during the rupture of a spreading drop arising from spontaneous liquid-liquid phase separation. The excess free energy released from phase separation is converted to the interfacial energy of the microdroplets and their kinetic energy. The resulting velocity of the propelling microdroplets reaches $\sim$ 160 $\mu$m/s at a maximum. The concentration gradients responsible for the formation and the initial spreading of the drop are intrinsically self-sustained through droplet dynamics along the wall surface and the induced liquid transport between the walls. We show that the phenomenon of self-propelling droplets is common for various unstable liquid mixtures even with high viscosity. Though a fully quantitative understanding of the rich and complex phenomena in the large space of control parameters is beyond the scope of this paper, we expect that the knowledge gained from this work will be valuable in designing new strategies for speeding up the transport in narrow spaces by leveraging the autonomous motions of microdroplets.

\section{Experimental section}
\subsection{Preparation of chemicals and substrates}
Solutions consisting of butyl paraben ($>$99 \%, Sigma-Aldrich) and ethanol (histological grade, Fisher Scientific) were prepared at mass ratio of 8/92, 10/90 or 15/85. Similar solutions were prepared using oleic acid (90\%, Alfa Aesar) or 1-octanol ($>$99 \%, Sigma-Aldrich). For experiments with increased viscosity, glycerol (Fisher Scientific) was added (see supporting information for table with detailed compositions). The viscosity of glycerol solutions with 30 wt\%, 50 wt\%, and 70 wt\% are 2.5, 6, and 25.5 cP, respectively \cite{segur1951}.

Hydrophobic Silicon (Si) substrates were prepared by coating with octadecyltrichlorosilane (OTS-Si) according to a previously reported method.\cite{zhang2008droplets} Briefly, Si substrates were cleaned with piranha solution (30\% $H_2O_2$ + 70\% $H_2SO_4$), de-ionized water (Mill-Q) and ethanol, respectively. Subsequently, the cleaned substrates were dried and incubated in a solution of OTS in hexane for 12 hr. Afterwards, the OTS coated Si wafers were sonicated for three times in OTS-free solvents for 10 min to remove loosely adsorbed OTS. All chemicals were purchased and used as received.


The interfacial tension of water-lean and water-rich subphases after phase separation was measured by using a Spinning drop tensiometer (SDT) (Kr\"{u}ss, Germany). The two subphases were made by adding water to a solution of oleic acid/water/ethanol (4.2/28.8/67 by vol\%) to drive phase separation.

\subsection{Experimental setup and visualization of propelling droplets}
The flow chamber was assembled by placing the Si substrate between a polycarbonate base and a cover glass sealed with a silicone rubber spacer with a thickness of 1 $mm$. The distance from the top cover glass to the substrate surface was 20-30 $\mu m$ after the chamber was assembled. The chamber was filled with the ternary solution from an inlet, followed by the injection of the displacing liquid (water, water/ethanol, or water/glycerol solutions) from one end of the side channel at a constant flow of 200 $\mu L/min$ with a syringe pump. The mixing of the ternary solution and the displacing liquid led to the droplets jetting phenomenon in the main microchannel.

The entire process of phase separation induced droplet propulsion was visualized using an optical microscope (Nikon H600L) equipped with a high speed camera (Photron Fastcam SA-Z) recording at 1000 frames-per-second (fps). 

To obtain the fluorescence data, we added trace of Nile Red (0.02 \%) to the binary solution (butyl paraben/ethanol = 10/90 wt/wt) with water as displacing liquid. The fluorescence intensity profiles around the hot spot was analyzed using self-written codes to convert the brightness in fluorescent images at different time points into normalized intensity. The composition map and intensity profiles were obtained along parallel lines at different locations. Also, the profile on the same location was tracked over a time period of 12 seconds.

A confocal microscope (SP5, Leica) was used to obtain an 3-dimensional image of self-propelled droplets after being pinned on the surface. The initial solution with the same concentration of Nile Red was prepared as the fluorescence experiment. A green HeNe laser was used to excite the fluorescent Nile Red dye. The contact angle of the pinned droplet was obtained from the 3D confocal image.

\subsection{Data analysis}
Droplets in all frames of a high-speed video were visualized and the positions of the droplets at different time were analyzed during the process from phase separation to droplet immobilization. Primary droplets spreading into a film and subsequent elongation into filaments, secondary droplets breaking off from the filament, and  self-propelling were tracked manually using ImageJ. The droplets were tracked until they stagnated or coalesced with neighboring droplets. The data were processed to obtain displacement values, velocities and locations within the fluid cell.

\subsection{Numerical simulation for diffusion of ethanol from solution A to solution B}
Numerical simulation of the receding oil-rich region was done via immersed boundary methods.\cite{rajat2005} Given the narrow thickness of the channel in the experiment setup, the two-dimensional diffusion equation


$$\frac{\partial c}{\partial t} = D\nabla^2c$$

was solved to obtain the temporal and spatial evolution of the concentration gradient of the solvents. Here $D$ is the diffusion coefficient taken as 2 $\times 10^{-9} m^2/s$. The oil phase with concentration of 1 (\textit{c} = 1) was initially restricted to an equilateral triangle area with size length $=$ 350 $\mu m$. The domain size was $L_{x} \times L_{y}$ = 2 $mm$ $\times$ 1 $mm$. 

At the wall of x = $L_{x}$, y = 0 and y = $L_{y}$, the concentration is set to 0. At the wall x = 0, the concentration obeys von Neumann boundary conditions,

$$\frac{\partial c}{\partial y} = 0$$ 
Note that the velocity field is not modeled, so in this simple approach $\vec{u}(\vec{x},t) \equiv{0}$, i.e. pure diffusion.

\subsection{Prediction of excess Gibbs free energy using UNIFAC}
The excess Gibbs free energies for butyl paraben-water-ethanol ternary mixture were predicted using the UNIQUAC Functional-group Activity Coefficients (UNIFAC) thermodynamic model\cite{fredenslund1975}. For all paraben concentrations, the tie lines in the ternary diagram were assumed to have the same ethanol composition in oil-rich and water-rich subphases as reported elsewhere\cite{yang2015phase}. The mole fraction of the water-rich subphase was taken to be the point where dilution line meets the phase envelope. The mole fraction of oil-rich subphase was estimated to be at the point on the opposite end of the tie line where it meets the phase envelope. For the composition of unstable mixture that would exist - if no phase separation had taken place - was calculated by setting the mass of water-rich subphase to be 100 times the mass of oil-rich subphase, based on the ratio of these two subphases along the tie lines in the phase diagram.  The group volume ($R_k$), surface area ($Q_k$) and interaction parameters ($a_{mn}$) used for UNIFAC are shown in Table 1 and 2, and temperature was taken as $T$ $=$ 298.15 $K$.

\section{Results and discussion}
\begin{figure}
	\begin{center}
		\includegraphics[trim={0cm 0cm 0cm 0cm}, clip, width=0.90\columnwidth]{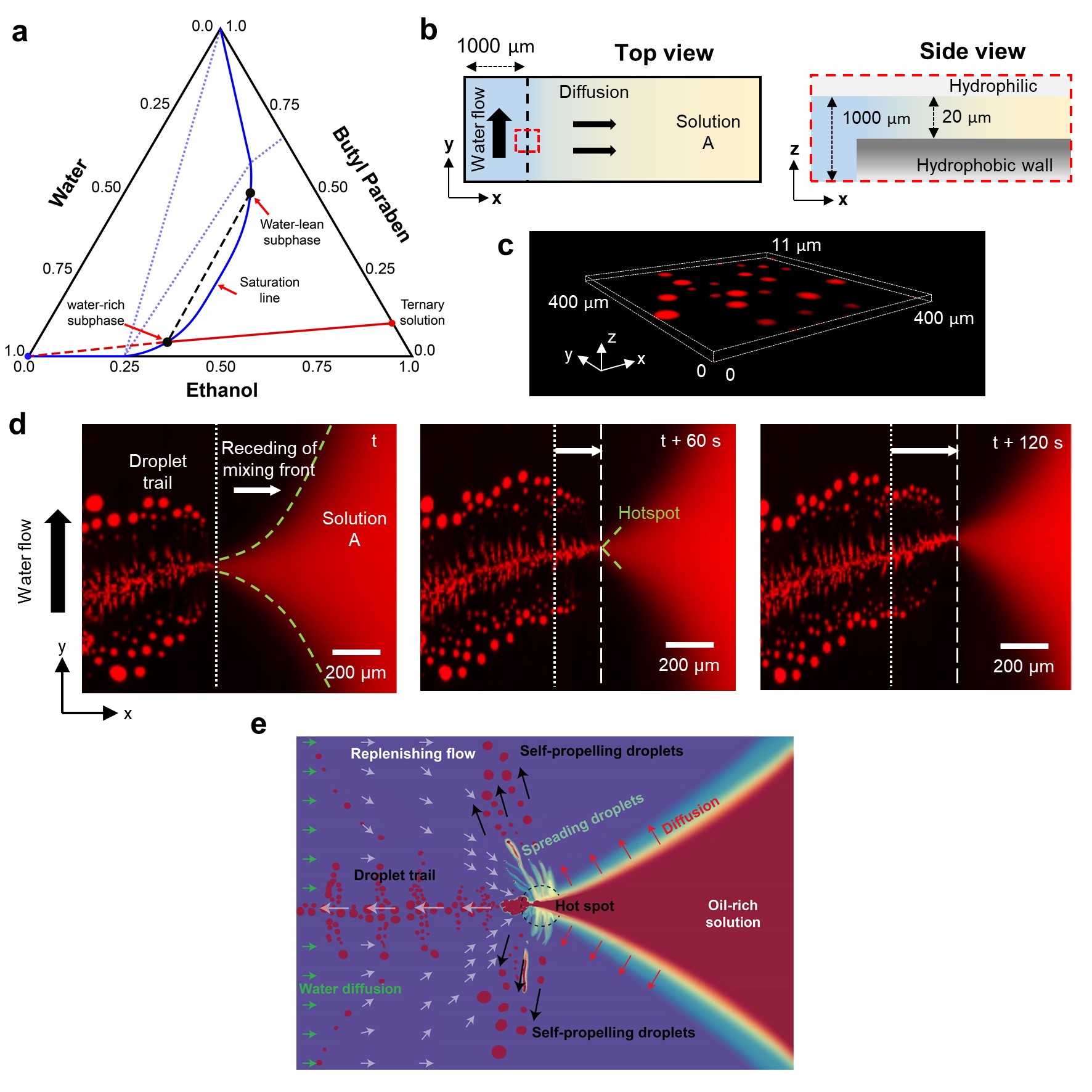}
		\captionsetup{font = small}
		\caption{Experimental configuration for self-propelling droplets and enhanced transport in a 2D chamber. \textbf{(a) }Ternary phase diagram for butyl paraben/ethanol/water system. Reproduced from Ref. \cite{yang2015phase} (Copyright 2015 Elsevier). \textbf{(b) }Sketch of the 2D chamber with top view and cross-section of the main and side channels. The surface of the bottom plate was hydrophobic while the top surface was hydrophilic. Solution A is initially encased in the microchamber with the composition as indicated in the solubility diagram. Water flows to the side of the encased solution and diffuses into it. The boundary between water and the oil solution moves from the side channel into the 2D main channel. The composition of the mixture is indicated by the dilution line (red dotted line in the phase diagram). An oil-rich subphase (red circle) and a water-rich subphase (blue star) form from liquid-liquid separation in response to oversaturation of paraben. \textbf{(c)}Confocal image showing the oil-rich subphase droplets formed on the hydrophobic wall of the microchamber. \textbf{(d) }Time-lapse fluorescence images show that the oil-rich subphase droplets are propelled along the bulk water flow while the mixing front moves toward the main channel. \textbf{(e)} Artistic sketch of the dynamical events occurring at a hot spot triggered by spontaneous phase separation.} 
		\label{illustration}
	\end{center}
\end{figure}

The first ternary solution we tested consisted of butyl paraben (a hydrophobic compound), a cosolvent ethanol, and a poor solvent water (cases 1 $\sim$ 3 in Table \ref{composition space}). Initially, the ternary solution filled both the microchamber (20 $\mu m$ in depth) and the side channel (1000 $\mu m$ in depth) as sketched in Figure \ref{illustration}. The top wall of the microchamber was a hydrophilic glass slide and the bottom wall, a smooth silicon substrate coated with a hydrophobic layer. Water was injected into the deep side channel to displace the ternary solution in the y-direction, and some of it diffused transversely into the microchamber in the x-direction.

A sequence of fluorescence images revealed a smooth boundary formed between the solution (doped with a dye) and water over time.  However, there were several triangular zones of solution protruding out from the smooth boundary into the water on the surface of the hydrophobic wall. At the hot spots of the triangle tips, microdroplet trails of several millimeters were left behind on the surface. From confocal microscopic imaging, the base radius of the microdroplets was found to range from $\sim$ 2 $\mu m$ to 25 $\mu m$, the contact angle was around $15^o-20^o$, and the maximal height was 0.5 $\mu$m to 7 $\mu$m above the surface.

Below we show that interestingly these trails consist of pinned microdroplets at the end of their propulsion. The motion of these self-propelling microdroplets is a key step in a cascade of dynamic events originating from liquid-liquid phase separation in 2D confinement as sketched in Figure \ref{illustration}e.

\subsection{Local concentration gradient and formation of self-propelling microdroplets}
\begin{figure}[htp]
	\begin{center}
		\vspace{-0.5 in}
		\includegraphics[trim={0cm 0cm 0cm 0cm}, clip, width=0.9\columnwidth]{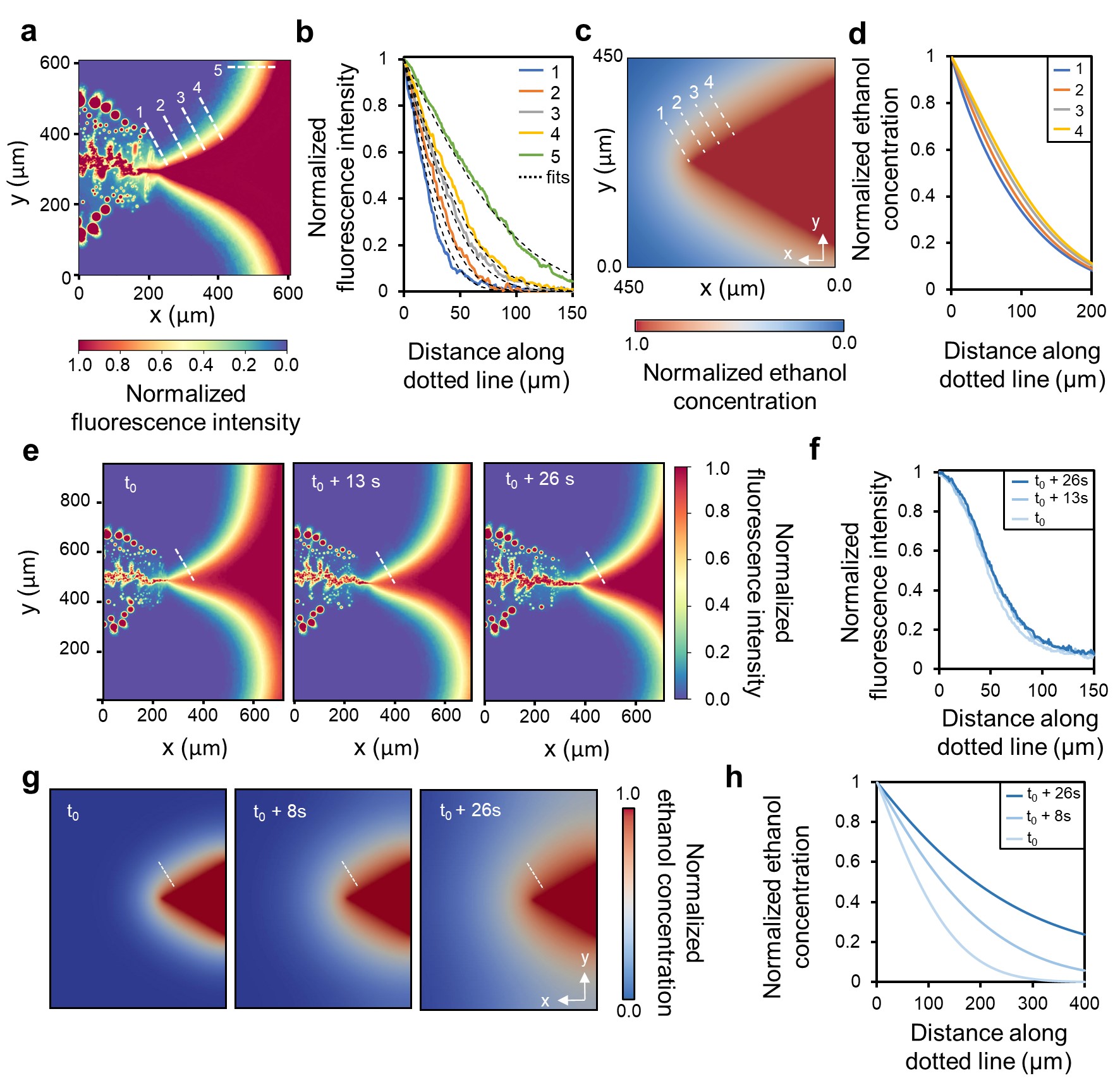}
		\captionsetup{font = small}
		\caption{\textbf{(a)} Intensity field of a receding mixing front at a fixed time obtained from fluorescence imaging. \textbf{(b)} Plot showing the intensity of fluorescence along lines 1 to 5 on the concentration field. The intensity along each line decays according to the error function, $I (x) \sim I_{0}e^{-(x/a)^2}$ where $I_{0}$ is taken as 1 and a is a constant. The fitted curves are shown as black dotted curves on the plot. The slope of intensity gradient becomes steeper as the location of the line in \textbf{(a)} approaches the tip (lines 5 to 1). This suggests that the concentration gradient is the sharpest at the hot spot where phase separation occurs. \textbf{(c)} Similar results are observed from a concentration field of ethanol at time t$_{0}$ = 4s obtained by numerical simulation. Here, the domain size is $L_{x} \times L_{y}$ = 450 $\mu m$ $\times$ 450 $\mu m$. \textbf{(d)} Concentration profiles along the lines 1 to 4 obtained from the concentration field. In agreement with experimental results, the slope for the concentration profile becomes steeper as the location of the line approaches the tip. \textbf{(e)} Time-lapse fluorescence images of intensity field over 26s and \textbf{(f)} intensity profile along the white dotted lines on the images. The intensity gradient at fixed location in the mixing front remains unchanged thanks to the replenishment of fresh water by recirculating flow. \textbf{(g)} Concentration field and \textbf{(h)} profiles on line 3 at various times obtained from numerical simulation. Unlike the experimental results, the stimulated concentration gradient smooths out over time.}
		\label{composition}
	\end{center}
\end{figure}

The spatial distribution of chemical composition was revealed by the fluorescence intensity of an oil-soluble dye (Nile red) doped in the ternary solution. The concentration ratio of ethanol to dye in the solution prior to phase separation was assumed to be constant as mixing with water dilutes them equally. Therefore, the fluorescence intensity indicates the concentration of both the dye and ethanol in the mixture. Figure \ref{composition}(a)(b) shows that at a given time, the ethanol concentration gradient across the mixing boundary becomes sharper near the tip of the triangular zone. The gradient is the sharpest at the hot spot and diminishes far away from the hot spot.

To mimic this key feature of the process, we simulated the diffusive mixing between the ternary solution and water in a triangular zone by solving the 2D diffusion equation, employing a finite difference code and the immerse boundary method \cite{rajat2005}. For details, we refer the reader to the method section. Figure \ref{composition}c shows the concentration field at $t = 4s$ after ethanol starts diffusing uniformly from the triangular zone into water. The concentration distribution of ethanol is consistent with the experimental measurements: a sharper concentration gradient at the tip of the triangle, i.e. the hot spot.

Diffusive mixing with water reduces both the concentration and the solubility of paraben in the ternary solution. While supersaturation of paraben can be realized and sustained under a limited set of conditions, more typically as composition reaches saturation at the hot spot, the mixture undergoes spontaneous liquid-liquid phase separation \cite{vitale2003}. A large fraction of the mixture ($\sim 99\%$) forms a water-rich subphase, while a small fraction forms droplets of a water-lean subphase (Figure \ref{illustration}a).

As the highest concentration gradient is located at the triangle tip, this is where microdroplets nucleate exclusively. The  concentration gradients along the smooth receding boundary are not sharp enough for the phase separation to initiate. The exact location and number of the triangular zones and the hot spots are possibly determined by the initial conditions of mixing, such as the influence of imperfection on the edge of the side channel on the formation of very early droplets\cite{lu2017}. Once formed, the triangular zone sustains along the receding boundary. The droplets appear brightest in the fluorescence images, because of enrichment of the hydrophobic dye in the water-lean droplets through nanoextraction \cite{lu2017,li2019functional}.

As noted above, the concentration profile in the simulations represents the effect from pure diffusion processes adjacent to the triangular zone. Differences between the experimental measurements and simulation results partly reflect effects beyond pure diffusion, for instance convective effect. The first difference is that the increase in the concentration gradient close to the hot spot is more pronounced in the experiments than in the simulations. The second difference lies in the trend in the temporal evolution of local concentration gradients. In Figure \ref{composition}e-h the concentration at the same location is plotted at time $t_{0}$ to $t_{0}+26s$ for experiment and the same time window for simulation. With time the simulated ethanol concentration gradient decayed by diffusion at a fixed location. On the other hand, the ethanol concentration gradient estimated from the composition map in the experiments was unchanged over a long period of time. Within the same distance from a hot spot, the concentration gradient of ethanol remained the same as the boundary recedes. The presence of such a local sharp and persistent gradient of ethanol clearly requires a pathway to supply water and to keep the mixture out of equilibrium at the hot spots. Below we show that this pathway is possibly through self-propelling microdroplets and the induced bulk flow. The ethanol concentration gradient in the y-direction is symmetric from the hot spot in both experimental and simulation results (Supporting Figure S1), which has important implications for the dynamics of droplet movement.

\subsection{Dynamics of self-propelling microdroplets}
\begin{figure}[htp]
	\begin{center}
		\includegraphics[trim={0cm 0cm 0cm 0cm}, clip, width=0.9\columnwidth,keepaspectratio]{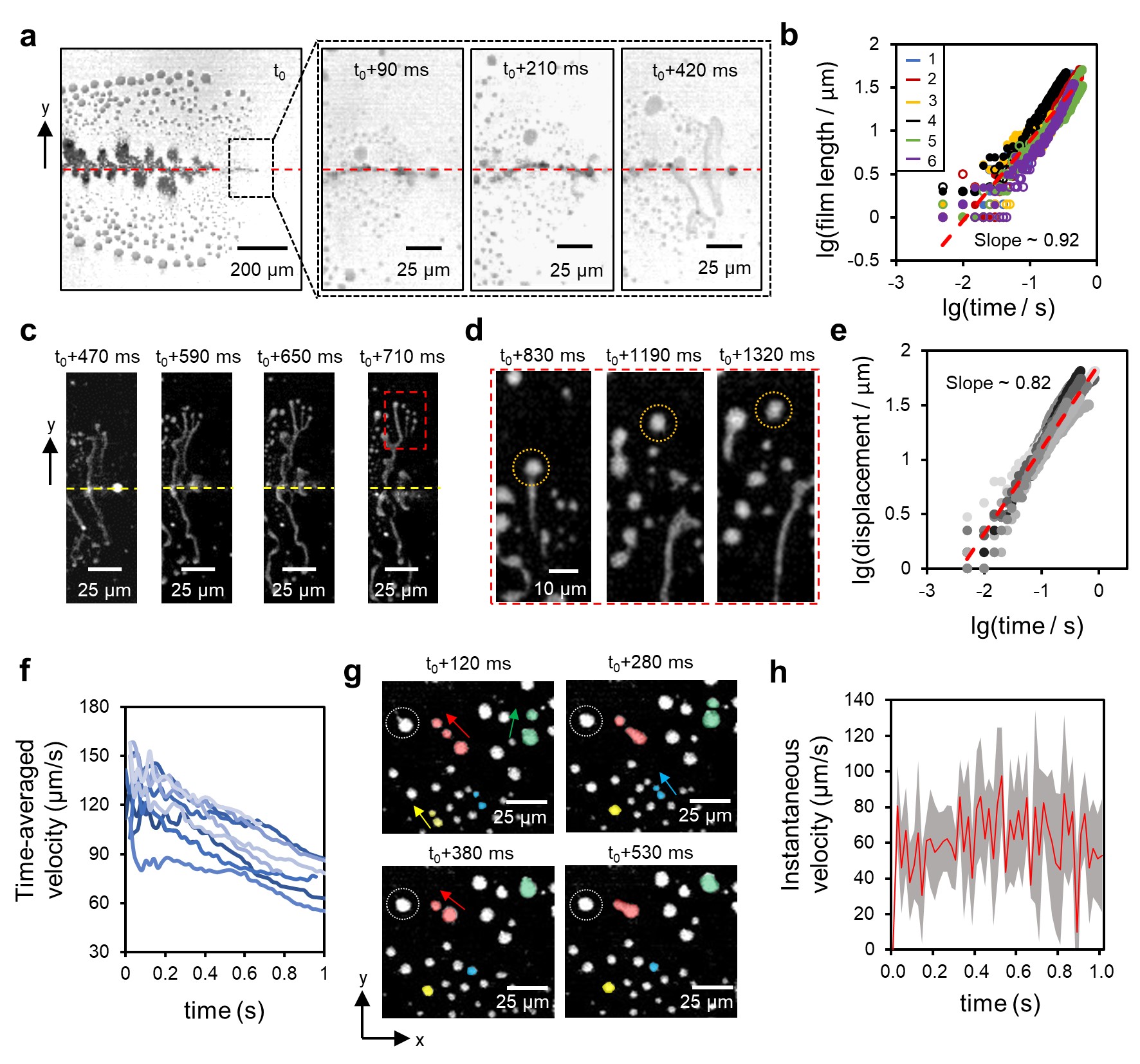}
		\captionsetup{font = small}
		\caption{Microscopic dynamics of the process. \textbf{(a)}  Time-lapse high-speed images showing water-lean subphase film spreading in $\pm$ y direction as a result of spontaneous phase separation at the hot spot. \textbf{(b)} Plot showing displacement vs. time for films spreading during 6 consecutive events. Filled and hollow dots represent the top and bottom half portions of the film with respect to the center indicated by the red dotted line in \textbf{(a)}. The dynamics of the film spreading is consistent with early dynamics of a drop spreading on a surface, i.e $x \sim t^{n}$, with $n \sim$ 1. Time-lapse high-speed images showing \textbf{(c)} the splitting of a film into several branches after sufficient spreading and \textbf{(d)} droplets being pinched off from a branch. \textbf{(e)} Plot showing displacement vs. time for 10 individual droplets pinched off from branches. The dynamics of droplets is similar to that of a spreading film, but with lower exponent, i.e. $x \sim t^{n}$ where $n \sim 0.82$. \textbf{(f)} Time-averaged velocities of droplets pinched off the branches. \textbf{(g)} Time-lapse images of droplets coalescing with each other after traveling a certain distance. Thanks to the drag from the surface, they eventually come to a stop.\textbf{(h)} Instantaneous point-to-point  velocities of self-propelled droplets tracked over 1 s. The red curve shows the mean data obtained from 10 self-propelled droplets and the gray colored area shows the maximum and minimum velocities at each instant in time.
		}
		\label{spreading2}
	\end{center}
\end{figure}

Our high speed images (Figure \ref{spreading2}, Movie S1) capture the rich and complex dynamics of droplets. At the start time ($t_{0}$), a thin line comprising several droplets forms at the tip of the trail that subsequently merges to form a larger drop with a diameter of 10 $\mu m$ at $t_{0} + 90$ ms. This drop becomes unstable at $t_{0} + 210$ ms, rapidly spreading a symmetrically from the hot spot under a stress from symmetrical distribution of local chemical compositions. Over a span of 480 ms, the average speed of the spreading front from six repeating events is similar in both +y and -y directions, with $(90 \pm 23)$ $\mu m /s$ and $(72 \pm 21)$ $\mu m /s$, respectively. The distance of the film front from the hot spot increases almost linearly with time ($\sim t^{0.92}$) as shown in the logarithmic plot in Figure \ref{spreading2}b.

The spreading film evolves into thin liquid filaments that then rupture at $t_{0} + 710$ ms, breaking up into shorter fragments and microdroplets with a diameter of (5 $\pm$ 1) $\mu m$. Microdroplets pinch off and jet out from the filaments, travelling at an initial velocity of $(73 \pm 46)$ $\mu m /s$, a similar velocity to that of the front of the stretching film. 

The film spreading and breakup of offspring microdroplets proceeds symmetrically in the y-direction, signifying symmetry in the stress acting on the water-lean subphase.
These self-propelling droplets continue to travel along the y-direction shown in Figure \ref{spreading2}d. Self-propelling droplets reach a maximum velocity of roughly 110 to 160 $\mu m/s$ (from 10 representative droplets) in the first tens of milliseconds. 

Self-propelling droplets slow down and evolve gradually to a crawling motion (Figure \ref{spreading2}g), due to resistance from both the surrounding liquid and the solid wall surface. The displacement of the microdroplets from the breakup point increases with time ($\sim t^{0.82}$). The crawling droplets are repeatedly stretched away from the hot spot and relaxed to become spherical cap (Movie S2), every 50 ms switching between pinning and depinning modes while moving away with an instant velocity fluctuating around 60 $\mu m/s$. This crawling motion of the droplets resembles a self-running droplet on a surface with wettability gradient \cite{varagnolo2013} or that of a light-driven droplet on a photoresponsive surface\cite{kavokine2016}, both driven by an  interfacial tension gradient. The crawling droplets coalesced with other slower or already-immobilized droplets and formed a larger droplet. More microdroplets coalesced on the way as the drop travels further until the droplets eventually reached the outer rim of the droplet trail and became immobilized. Some droplets barely moved after nucleation, forming the central line of the trail.

The above results show that on the two rims, microdroplets come to the end of the motion of self-propulsion. The width of the droplet trail is determined by the longest distance travelled by the microdroplets from the hot spot. Next we analyze the local chemical composition in space and time to reveal why the phase separation only takes place at the hot spot, and why the droplets move away from the hot spot.

\subsection{Induced directional flow in the bulk between the walls}
\begin{figure}[htp]
	\begin{center}
		\includegraphics[trim={0cm 0cm 0cm 0cm}, clip, width=0.9\columnwidth]{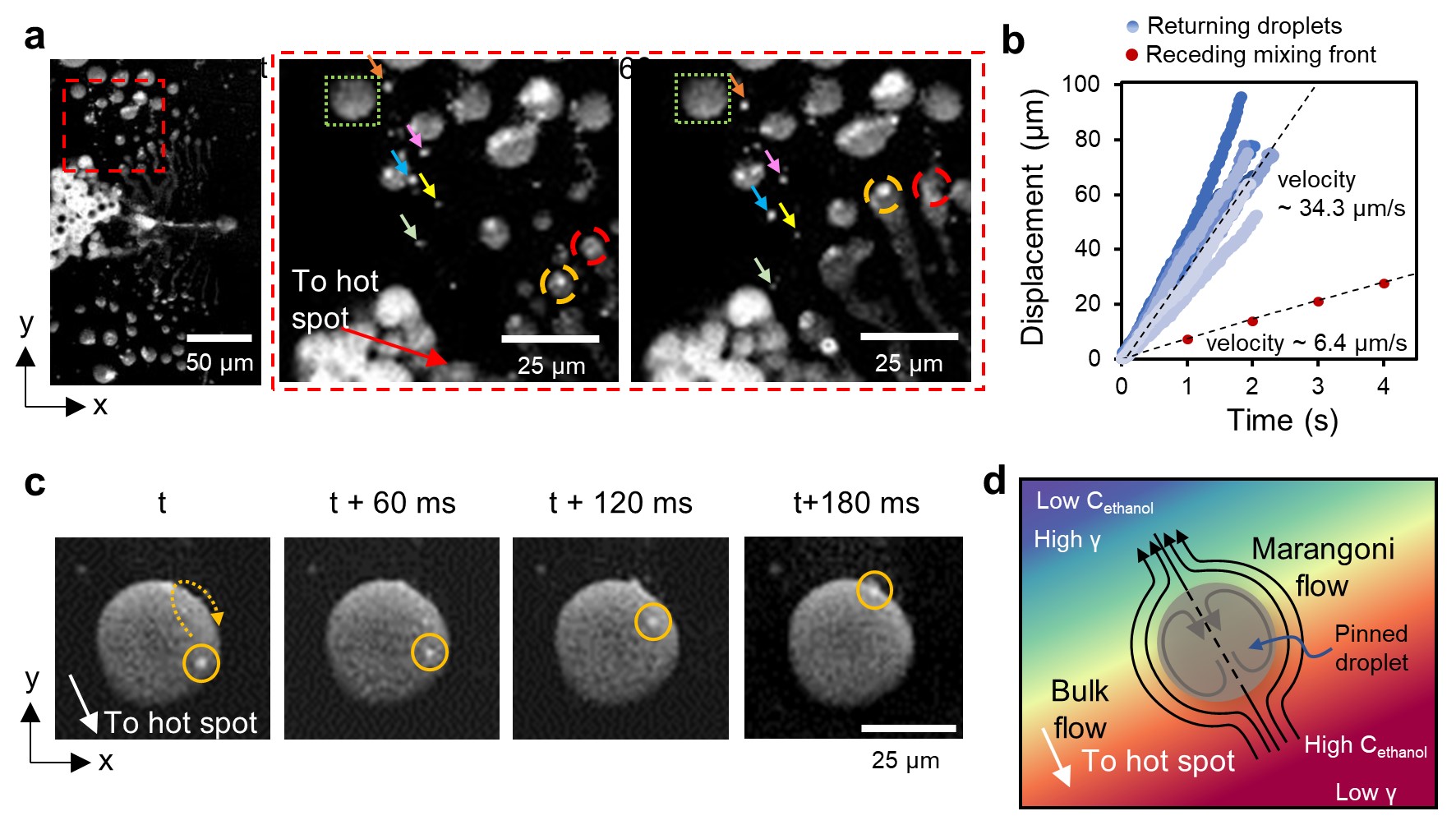}
		\captionsetup{font = small}
		\caption{\textbf{(a)} High-speed time-lapse images demonstrating the presence of replenishing flows visualized by tiny oil droplets traveling towards the hot spot. \textbf{(b)} Plot showing the displacement vs. time of returning oil droplets and the receding mixing front. The timescales of the two are vastly different. The replenishing flow travels at $\sim$ 34 $\mu m/s$ whereas the receding mixing front only travels at $\sim$ 6.4 $\mu m/s$.  \textbf{(c)} The motion of tiny oil droplets in the replenishing flow is influenced by the Marangoni flow around large pinned droplets occurring due to the ethanol concentration gradient. As a result, once tiny oil droplets are close enough to the large droplets, the competition between replenishing flow and Marangoni flow drives the tiny oil droplets in a circulatory motion around the large droplets as captured by high-speed time-lapse images. \textbf{(d)} Schematics showing the presence of bulk flow and Marangoni flow around a large pinned droplet.}
		\label{flow}
	\end{center}
\end{figure}

In the direction opposite to that of the self-propelling droplets, we observed a slower flow in the bulk between the walls that is induced by self-propelling droplets due to mass conservation in the confined space, see Figure \ref{flow}. Across an area of a semi-circular sector towards the hot spot, this bulk flow is manifested by the trajectory of very small oil droplets with a radius of $\sim$ 1 $\mu m$ dispersed in the solution that act as tracers. By tracking these droplet tracers suspended in the flow, the average velocity of the flow was determined to be 34 $\mu m/s$ at the distance of $\sim$ 150 $\mu m$ from the hot spot. Closer to the hot spot, the flow was faster, reaching an average speed of (46 $\pm$ 15) $\mu m/s$. After reaching the hot spot, the flow converged to a single stream, heading toward the side channel along the droplet trail. This bulk flow brings fresh water to the hot spot continuously, maintaining the sharp concentration gradient at the hot spot.

The pinned microdroplets on the wall surface altered the local velocity profile of the replenishing flow. When the droplet tracers in the flow approached a pinned microdroplet, they were trapped by a local flow near the microdroplet surface. The revolving motion of the tracers is attributed to the concentration gradient of solvent constituent around the droplets. In the proximity to the pinned droplet, the Marangoni stress is strong enough to overcome the directional flow in the bulk and to spin the tracers.  Although the ethanol concentration was too low to be reflected in the fluorescence intensity, the variation of ethanol concentration in space was sufficient to induce a gradient in the interfacial tension of the droplet surface and a local Marangoni flow adjacent to the droplet.

\subsection{Generality of self-propelling microdroplets}
\begin{figure}[htp]
	\begin{center}
		\includegraphics[trim={0cm 0cm 0cm 0cm}, clip, width=0.9\columnwidth]{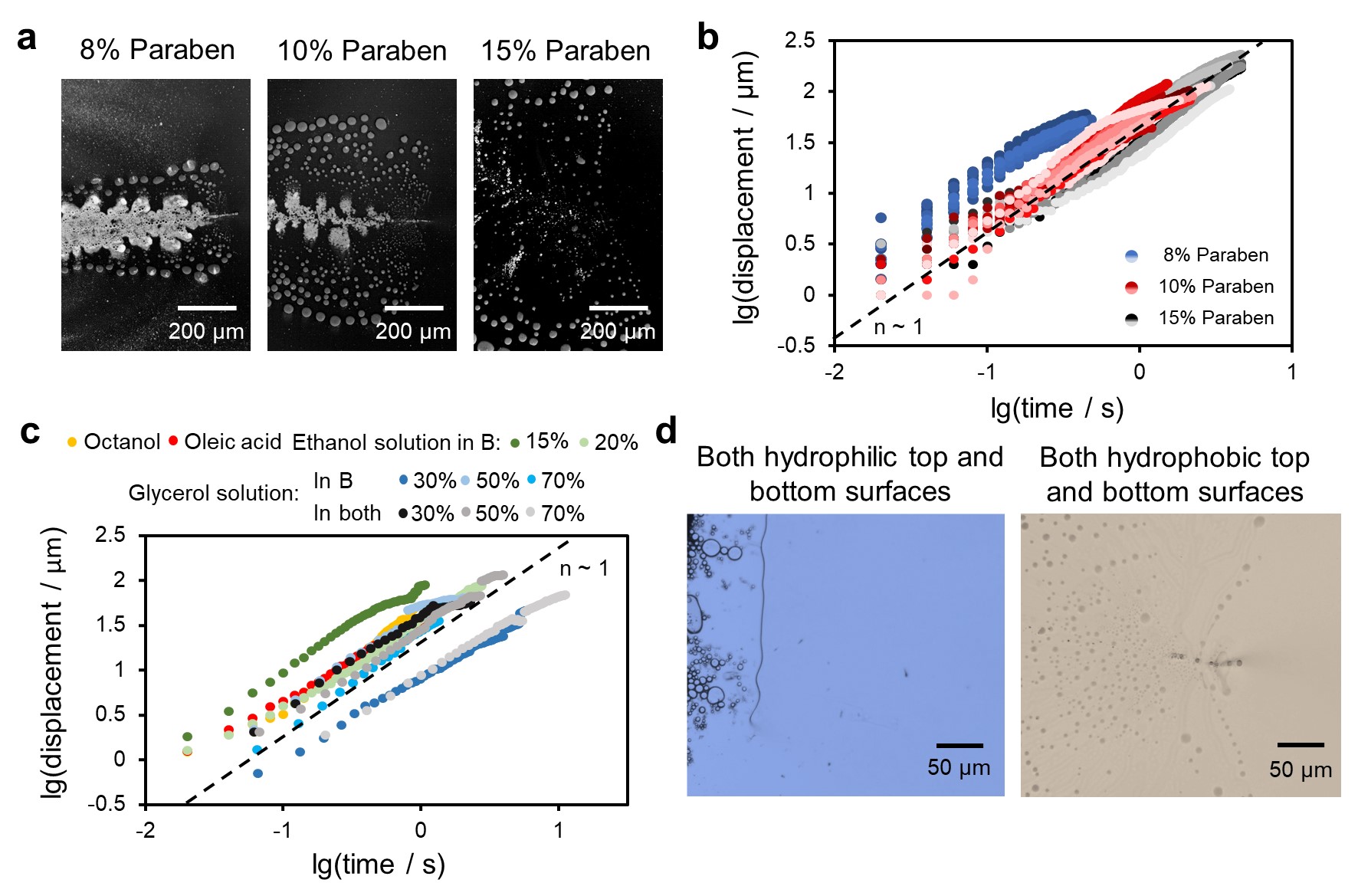}
		\captionsetup{font = small}
		\caption{\textbf{(a)} The self-propulsion of droplets driven by phase separation occurs regardless of initial paraben concentration as shown by the time-lapse high-speed images. \textbf{(b)} Plot showing the displacement vs time of 10 individual droplets for initial paraben concentrations of 8\%, 10\% and 15\%. All data follow a similar trend as that of a spreading film, i.e. x $\sim$ t$^{n}$ where n $\sim$ 1. \textbf{(c)} The dynamics shown in this work also holds for various oil types, compositions of Solution B, and viscosities of both Solutions A and B. Again, the data follow a similar trend. \textbf{(d)} Surface wettability is critical for the phenomena shown in this work. Optical images of droplet formation in a chamber with both top and bottom surfaces hydrophilic (left) and hydrophobic (right).}
		\label{universality}
	\end{center}
\end{figure}

We now further explored the generality of the self-propelling droplets with a parametric investigation. The parameters included the chemical composition, the type, and viscosity of the ternary mixtures, and the wettability of the wall surfaces, as summarized in Table \ref{composition space}. The directional flow and self-propelling droplets occur in a certain range of chemical composition, viscosities and oil types on a hydrophobic wall surface. The velocity of self-propelling droplets was related to the initial chemical composition of the ternary solution (Figure \ref{universality}). At a given concentration of paraben, adding 10 \% - 20 \% ethanol in water as the displacing liquid does not change the speed of the self-propelling droplets but reduces the number of locations of phase separation. In the range of 8 \% - 15 \%, the droplets travel a shorter distance but move faster at a low concentration of paraben. The results show that tuning the chemical composition is another way to vary the velocity of self-propelling droplets.

The same dynamics are observed in the solutions with higher viscosity (2.5 cP to 25.5 cP) and in various ternary mixtures (such as 1-octanol or oleic acid in isopropanol and water). How far a droplet travels before coalescence or how fast a droplet moves was found to vary in each case. However, linear droplet displacement $l$ with time ($l \sim t $) was observed for all cases. The results suggest the mechanism for self-propelling microdroplets is general for liquid-liquid phase separation of ternary systems.

The wettability of the wall surface plays an important role in self-propelling microdroplets. In a microchamber with both walls being hydrophobic, the velocity of self-propelling droplets can reach $(111 \pm 6)$ $\mu m/s$ on average, almost twice the speed of droplets where only the bottom surface is hydrophobic. The droplet trail is absent as all droplets move away from the hot spot. For a microchamber with both walls hydrophilic, all droplets are pinned on surfaces and there is no enhanced fluid motion.

\subsection{Mechanisms for self-propelling droplets and directional flow}
In the proposed mechanism illustrated in Figure \ref{illustration}e, liquid-liquid phase separation are induced bulk flow are linked in a self-sustained cycle. The sharp concentration gradient leads to local liquid-liquid phase separation at the hot spot where the concentration of ethanol fades away while water concentration increases. The interfacial tension is negligible between the water-rich subphase and water-lean droplets, but is substantial between water and droplets.  For instance, the interfacial tension is only $\sim$ 0.2 $mN/m$ between two equilibrium subphases formed from phase separation in the mixture of water, ethanol and oleic acid. In contrast, the interfacial tension reaches $\sim$ 11 $mN/m$ for an oleic acid droplet and water \cite{kallio2009}. A positive interfacial tension gradient away from the hot spot exerts a Marangoni stress on the free surface of the droplet, leading to symmetrical spreading along the y-direction. Under stress the spreading film eventually becomes unstable, breaking up into smaller droplets.

The driving force for droplet spreading in our experiments is similar to film spreading enhanced by a concentration gradient of a surfactant or an organic solvent in the formation of Marangoni flowers \cite{wodlei2018}, or spreading on a liquid surface driven by solvent evaporation \cite{durey2018,kim2017-stone}. In all cases a small volume of liquid is exposed to a concentration gradient of the external constituents.

For a given droplet volume, the distance of the film front ($l$) under the stress is linear in time, i.e. $l \sim t$, completely different from drop spreading governed by capillary force according to Tanner's law ($l \sim t^{0.1}$).  This linear relationship holds for the temporal evolution of the spreading front, and initial movement of the self-propelling droplets after break-off from the film in Figures \ref{spreading2} and \ref{universality}.  The linear increase with time is also reported in spreading of aqueous surfactant drop on surfaces, enhanced by the Marangoni stress from the concentration gradient of a surfactant \cite{rafai2002}. The balance between the Marangoni stress and viscous stress on the droplets leads to the linear dependence $l \sim t$, as reported previously \cite{rafai2002}.

Further away from the hot spot, the gradient in ethanol concentration smoothens out, thus less stress propels the droplets. At the rim of the droplet trail, the stress from $\Delta \gamma$ is balanced by the viscous drag force from the surrounding fluid (now mostly water) and the friction from the wall surface. Finally, due to the viscous drag and frictional forces, the self-propelling droplet becomes pinned on the surface. The effect from the wall wettability on self-propelling droplets is evident as no self-propulsion takes place on a hydrophilic surface where the droplet cannot overcome the large static friction force and displace the external water-rich liquid on the surface \cite{gao2018}.

The induced flow enhances the sharp concentration gradient at the hot spot for further phase separation. At the same time, the mass loss at the hot spot of the self-propelling droplets and the outflux along the droplet trail is compensated by influx of the directional flow to the hot spot.  In this way, a self-sustained cycle completes and repeats itself from liquid-liquid phase separation at the hot spot.

To summarize, we note that the cycle consists of several interconnected steps on different time and length scales. The diffusive boundary moves away from the side channel for several 10$^{2}$ s in time and over centimeters in length scale. The formation and propulsion of droplets occurs in 10$^{-3}$ s over micrometers in distance. The induced flow travels over centimeters to reach the hot spot, several orders of magnitude faster than the receding boundary. This replenishing flow is 2-3 times faster than the diffusiophoretic transport induced by chemical gradient or catalytic reactions reported in the literature \cite{gao2014catalytic,hernandez2013}. Such enhanced transport may have important implications for mass transfer involving liquid-liquid phase separation in confined spaces.

We now focus on the source of energy input to the cycle. Spontaneous phase separation of the unstable mixture releases excessive Gibbs free energy. We estimated the excess Gibbs free energies of the ternary system in the states before and after phase separation using the UNIFAC model. The model enables prediction of activity coefficients based on the interaction of functional groups found in each component in a mixture.\cite{fredenslund1975} 
Using UNIFAC, the $\Delta G^{E}$ at initial oil concentration is estimated to be 1.89 $J/mol$ at 8 \% paraben, and decreases to 1.66 $J/mol$ at 10 \%, and further to 1.19 $J/mol$ at 15 \% (Table \ref{energies}).

A substantial portion of the excessive free energy from the phase separation is converted to interfacial energies in forming three interfaces: 1) the solid-liquid interface between water-rich subphase and substrate, 2) the solid-liquid interface between water-lean subphase and substrate, and 3) the liquid-liquid interface between the two subphases. The interfacial energies gained by the system from forming the three interfaces can be estimated by the difference between the interfacial energies between the unstable mixture and the substrate before and after phase separation, and the interfacial energies between the droplet and the surrounding liquid (Supporting Information). The total energy gain from the interfaces is approximately 0.5 $J/mol$, on the same order of magnitude as excessive free energy from phase separation estimated from UNIFAC.

We take different oil concentrations as the example to estimate whether the energy supplied from phase separation is sufficient to drive self-propelling droplets. The kinetic energy of droplets were calculated by $E_k \sim \frac{1}{2}mv^{2}$ where $m$ is the mass and $v$ is the velocity. The mass of each droplet was obtained using a density estimated based on the water-lean subphase composition and the volume of the droplet. The kinetic energy of autonomous droplets listed in Table \ref{energies} and is on the same order of magnitude as that for droplets driven by other means such as surface chemical gradient\cite{hernandez2013} or catalytic motor\cite{gao2014catalytic}. The kinetic energy of propelling droplets accounts for only $\sim$ 1 millionth of the total amount of the free energy released from phase separation at given chemical composition. Such low efficiency in the energy conversion to kinetic energy is an expected result, considering the large proportion of energy converted to interfacial energy. 

\section{Conclusions}
We show that microdroplets from liquid-liquid phase separation self propel and induce a fast flow in a quasi-2D microchamber.   The microdroplets form spontaneously from diffusive mixing of a ternary mixture and a poor solvent, spread rapidly, and travel across the wall surface. The self-propulsion of the droplets induces a replenishing flow to the location of phase separation, which sustains the chemical concentration gradients in further phase separation. Our thermodynamic estimation shows that excessive free energy from the phase separation is more than sufficient to account for the kinetic energy of the propelling droplets. The phenomena studied in this work may have significant implications for a wide range of practical processes, as they are general, taking place in liquid mixtures of different concentrations, solution viscosities and solvent combination.  Our findings highlight the possibility of fast and smart fluid transport by controlling phase separation behaviors in confined spaces, which is of major interest for extraction of liquids from porous media, separation by membrane technology, phase transfer catalysis, or biphasic chemical reactions.  

\section*{Declaration of Competing Interests}
The authors declare that they have no known competing financial interests or personal relationships that could have appeared to influence the work reported in this paper.

\section*{Acknowledgement}
The project is supported by the Natural Science and Engineering Research Council of Canada (NSERC) and Future Energy Systems (Canada First Research Excellence Fund). This research was undertaken, in part, thanks to funding from the Canada Research Chairs program. DL acknowledges the support from the European Research Council - Advanced Grant under Project Number 740479 and from the European Research Council - Proof-of-Concept Grant under Project Number 862032.

\bibliography{ref_cej}

\newpage

\newpage

\newpage

\newpage

\end{doublespace}	
\end{document}


\begin{frontmatter}

\title{Supporting information}

\author[1,2]{Xuehua Zhang\corref{cor1}}
\ead{xuehua.zhang@ualberta.ca}
\author[1,2]{Jae Bem You\fnref{fn1}}
\author[1]{Gilmar F. Arends\fnref{fn1}}
\author[1]{Jiasheng Qian\fnref{fn1}}
\author[2]{Yibo Chen}
\author[2,3]{Detlef Lohse}
\author[1]{John M. Shaw\corref{cor2}}
\ead{jmshaw@ualberta.ca}
\address[1]{Department of Chemical and Materials Engineering, University of Alberta, Alberta T6G 1H9, Canada}
\address[2]{Physics of Fluids Group, Max Planck Center Twente for Complex Fluid Dynamics, JM Burgers Center for Fluid Dynamics, Mesa+, Department of Science and Technology, University of Twente, Enschede 7522 NB, The Netherlands}
\address[3]{Max Planck Institute for Dynamics and Self-Organization, 37077 G\"{o}ttingen, Germany}

\fntext[fn1]{These authors contributed equally to this work.}
\cortext[cor1]{Corresponding author}
\cortext[cor2]{Corresponding author}
		
\end{frontmatter}

\section*{Estimation of interfacial energies}
Upon phase separation, a significant portion of the Gibbs free energy is used up in forming three interfaces: 1) solid-liquid interface between water-rich subphase and substrate, 2) solid-liquid interface between water-lean subphase and substrate, and 3) liquid-liquid interface between the two subphases. The energy required to form the three interfaces can be estimated by the difference between the interfacial energies between the mixture and the substrate before and after phase separation. Prior to phase separation, the interfacial energy between the unstable ternary mixture and the substrate ($E_{1}$)  is given by the product of interfacial tension ($\gamma_{ternary-S}$) with contact area ($A$),

$$E_{1} = \gamma_{ternary-S}A$$
Once phase separation takes place and the three interfaces form, the total interfacial energy ($E_{2}$) becomes the sum of the product of interfacial tensions and contact areas:

$$E_{2} = \gamma_{L1-S}A_{1} + \gamma_{L2-S}A_{2} + \gamma_{L1-L2}A_{interface}$$
$L1$, $L2$, and $S$ denote water-lean subphase, water-rich subphase and substrate, respectively. The $\gamma$ denote the interfacial tensions between respective interfaces. $A_{1}$ and $A_{2}$ denote the contact area of water-lean and water-rich subphases with the substrate, respectively, such that $A = A_{1} + A_{2}$. $A_{interface}$ is the interfacial area between the two subphases (See Figure \ref{energy}). The energy $E$ required to form the three interfaces is then the difference between the two states $E = E_{2} - E_{1}$.

Because the composition of each subphase is not exactly known, here we estimate the interfacial energy for the extreme case of an oil droplet (i.e. 1-octanol) surrounded by water. Therefore, in the following discussions, octanol and water are taken as $L_{1}$ and $L_{2}$, respectively.

With this assumption, $\gamma_{L_{1}-S}$ and $\gamma_{L_{2}-S}$ can be obtained using the Young's equation as shown below (see also Figure S2):

$$\gamma_{S} = \gamma_{L_{1}}cos(\theta) + \gamma_{L_{1}-S} = \gamma_{L_{2}}cos(\theta) + \gamma_{L_{2}-S}$$

where the surface tensions of octanol ($\gamma_{L_{1}}$) and water in air ($\gamma_{L_{2}}$), and the surface tension of OTS-coated substrate ($\gamma_{S}$) in air are known from literature. The static contact angle ($\theta$) of each liquid on the substrate with air as the surrounding is $\theta_{octanol} \sim$ 42 $^{o}$ and $\theta_{water} \sim$ 110 $^{o}$. Table \ref{tensions} lists the values of surface tension as well as solid-liquid interfacial tensions for each liquid. The interfacial tension between the liquids is also known from literature.

Once the interfacial tensions are known, the total energy can be calculated by multiplying the $\gamma$ values with the contact area. Here, the volume of surrounding phase (i.e. water) is estimated to be 100 times that of the oil, based on the ratio of the two subphases from phase separation along a tie line, see Figure 1a and the reference \cite{yang2015phase}. Thus, with an oil droplet of 2.5 $\mu m$ in radius, the contact areas with the substrate for both oil and water are $A_{1} \sim$ 1.96 $\times$ 10$^{-11}$ $m^{2}$ and $A_{2} \sim$ 1.34  $\times$ 10$^{-10}$ $m^{2}$, respectively. The interfacial area between the two liquids is $A_{interface}$ $\sim $ 5.4 $\times$ 10$^{-12}$ $m^{2}$. 

With these values, the total energy required for forming the three interfaces is $E = E_{2} - E_{1} \sim$ 0.49 $J/mol$. The energy approximated here is likely to be higher than the actual interfacial energy since the experimental value for interfacial tension between the subphases is very small (i.e. $\sim$0.2 $mN/m$ for oleic acid, ethanol, water mixture). However, this analysis demonstrates that most of the energy from phase separation is spent for the formation of new interfaces.

\setcounter{table}{0}
\makeatletter 
\renewcommand{\thetable}{S\arabic{table}}

\newpage
\begin{table}[ht]
	\captionsetup{font = {small}}
	\caption{Combinations of solutions and wettability of wall surfaces}
	\centering
	\resizebox{15cm}{!}{
	\begin{tabular}{|c|c|c|c|c|}
	    \hline
		\ & Solution composition (wt\%) & Displacing liquid (wt\%) & Top wall & Bottom wall\\
		\hline
		1 & Paraben (8) : Ethanol (92) & \multirow{3}{*}{Water (100)} & & \\
		2 & Paraben (10) : Ethanol (90) & & Hydrophilic & Hydrophobic \\
		3 & Paraben (15) : Ethanol (85) & & & \\
		\hline
		4 & \multirow{2}{*}{Paraben (10) : Ethanol (90)}& \multirow{2}{*}{Water (100)}& Hydrophilic & Hydrophilic \\
		5 & & & Hydrophobic & Hydrophobic \\
		\hline
		6 & \multirow{2}{*}{Paraben (10) : Ethanol (90)}& Ethanol (15): Water (85) & \multirow{2}{*}{Hydrophilic} & \multirow{2}{*}{Hydrophobic}\\
		7 & & Ethanol (20) : Water (80) & & \\
		\hline
		8 & \multirow{3}{*}{Paraben (10) : Ethanol (90)} & Glycerol (30) : Water (70) & \multirow{3}{*}{Hydrophilic} & \multirow{3}{*}{Hydrophobic}\\
		9 & & Glycerol (50) : Water (50) & &\\
		10 & & Glycerol (70) : Water (30) & &\\
		\hline
		11 & Paraben (10) : Ethanol (63) : Glycerol (27) & Glycerol (30) : Water (70) & \multirow{3}{*}{Hydrophilic} & \multirow{3}{*}{Hydrophobic}\\
		12 & Paraben (10) : Ethanol (45) : Glycerol (45) & Glycerol (50) : Water (50) & & \\
		13 & Paraben (10) : Ethanol (27) : Glycerol (63) & Glycerol (70) : Water (30) & & \\
		\hline
		14 & Oleic acid (4.2) : Ethanol (67) : Water (28.8) & \multirow{2}{*}{Water (100)} &  \multirow{2}{*}{Hydrophilic} & \multirow{2}{*}{Hydrophobic}\\
		15 & Octanol (8) : Ethanol (47) : Water (45) & & & \\
		\hline
	\end{tabular}}
	\label{composition space}
\end{table}

\newpage
\begin{table}[ht]
\captionsetup{font = {small}}
\caption{Characteristic group volume ($R_k$) and surface area ($Q_k$) parameters for each chemical group in our ternary systems. Here A represents an aromatic group. Data are taken from Dortmund Data Bank (DDB)\cite{DDBST}}
\centering
\begin{tabular}{|c|c|c|}
\hline
\ & $R_k$ & $Q_k$ \\
\hline
$CH_3$ & 0.9011 & 0.8480 \\
\hline
$CH_2$ & 0.6744 & 0.5400 \\
\hline
$COO$ & 1.3800 & 1.2000 \\
\hline
$ACOH$ & 0.8952 & 0.6800 \\
\hline
$ACH$ & 0.5313 & 0.4000 \\
\hline
$AC$ & 0.3652 & 0.1200 \\
\hline
$OH$ & 1.0000 & 1.2000 \\
\hline
$H_2O$ & 0.9200 & 1.4000 \\   
\hline
\end{tabular}
\end{table}

\newpage
\begin{table}[ht]
	\captionsetup{font = {small}}
	\caption{Group interaction parameters ($a_{mn}$) for the functional groups found in the system.  Here A represents an aromatic group. Taken from Dortmund Data Bank (DDB)\cite{DDBST} 
	\\}
	\centering
	\begin{tabular}{|c|c|c|c|c|c|c|c|c|}
	    \hline
		\ & $CH_3$ & $CH_2$ & $COO$ & $ACOH$ & $ACH$ & $AC$ & $OH$ & $H_2O$ \\
		\hline
		$CH_3$ & 0 & 0 & 387.1 & 1333 & 61.13 & 61.13 & 986.5 & 1318 \\
		\hline
		$CH_2$ & 0 & 0 & 387.1 & 1333 & 61.13 & 61.13 & 986.5 & 1318 \\
		\hline
		$COO$ & 529 & 529 & 0 & -167.3 & 317.6 & 317.6 & 88.63 & 284.4 \\
		\hline
		$ACOH$ & 275.8 & 275.8 & -494.2 & 0 & 25.34 & 25.34 & -451.6 & -601.8 \\
		\hline
		$ACH$ & -11.12 & -11.12 & 103.5 & 1329 & 0 & 0 & 636.1 & 903.8 \\
		\hline
		$AC$ & -11.12 & -11.12 & 103.5 & 1329 & 0 & 0 & 636.1 & 903.8 \\
		\hline
		$OH$ & 156.4 & 156.4 & 190.3 & -259.7 & 89.6 & 89.6 & 0 & 353.5 \\
		\hline
		$H_2O$ & 300 & 300 & -197.5 & 324.5 & 362.3 & 362.3 & -229.1 & 0 \\
		\hline
	\end{tabular}
\end{table}

\begin{table}[ht]
	\captionsetup{font = small}
	\caption{Surface and interfacial tensions of octanol ($L_{1}$), water ($L_{2}$) and OTS-Si substrate ($S$)}
	\centering
	\begin{tabular}{|c|c|c|}
		\hline
		& mN/m & \\
		\hline
		$\gamma_{S}$ & 20.7 &  ref. \cite{kulkarni2005}\\
		\hline
		$\gamma_{L_{1}}$ & 27.5 &  ref. \cite{yu2019}\\
		\hline
		$\gamma_{L_{2}}$ & 72 & \\
		\hline
		$\gamma_{L_{1}-L_{2}}$ & 8.4 & ref. \cite{yu2019}\\
		\hline
		$\gamma_{L_{1}-S}$ & 0.26 & calculated\\
		\hline
		$\gamma_{L_{2}-S}$ & 45.3 & calculated\\
		\hline
	\end{tabular}
	\label{tensions}
\end{table}

\newpage
\begin{table}[ht]
	\captionsetup{font = {small}}
	\caption{Energies associated in the process: Excess Gibbs free energy during phase separation ($\Delta G_{exc}$) from UNIFAC calculation and average kinetic energy of self-propelled droplets ($E_{kinetic}$)
		\\}
	\centering
	\begin{tabular}{|c|c|c|c|}
		\hline 
		Paraben concentration & $\Delta G$ (J/mol) & E$_{kinetic}$ (10$^{-11}$J/mol)\\
		\hline
		8\% & 1.89 &  45.3 $\pm$ 13.3\\
		\hline
		10\% & 1.66  & 8.93 $\pm$ 4.75 \\
		\hline
		15\% & 1.19 & 6.47 $\pm$ 1.0 \\
		\hline
	\end{tabular}
	\label{energies}
\end{table}

\setcounter{figure}{0}
\makeatletter 
\renewcommand{\thefigure}{S\arabic{figure}}

\newpage
\begin{figure}[ht]
    \centering
    \includegraphics{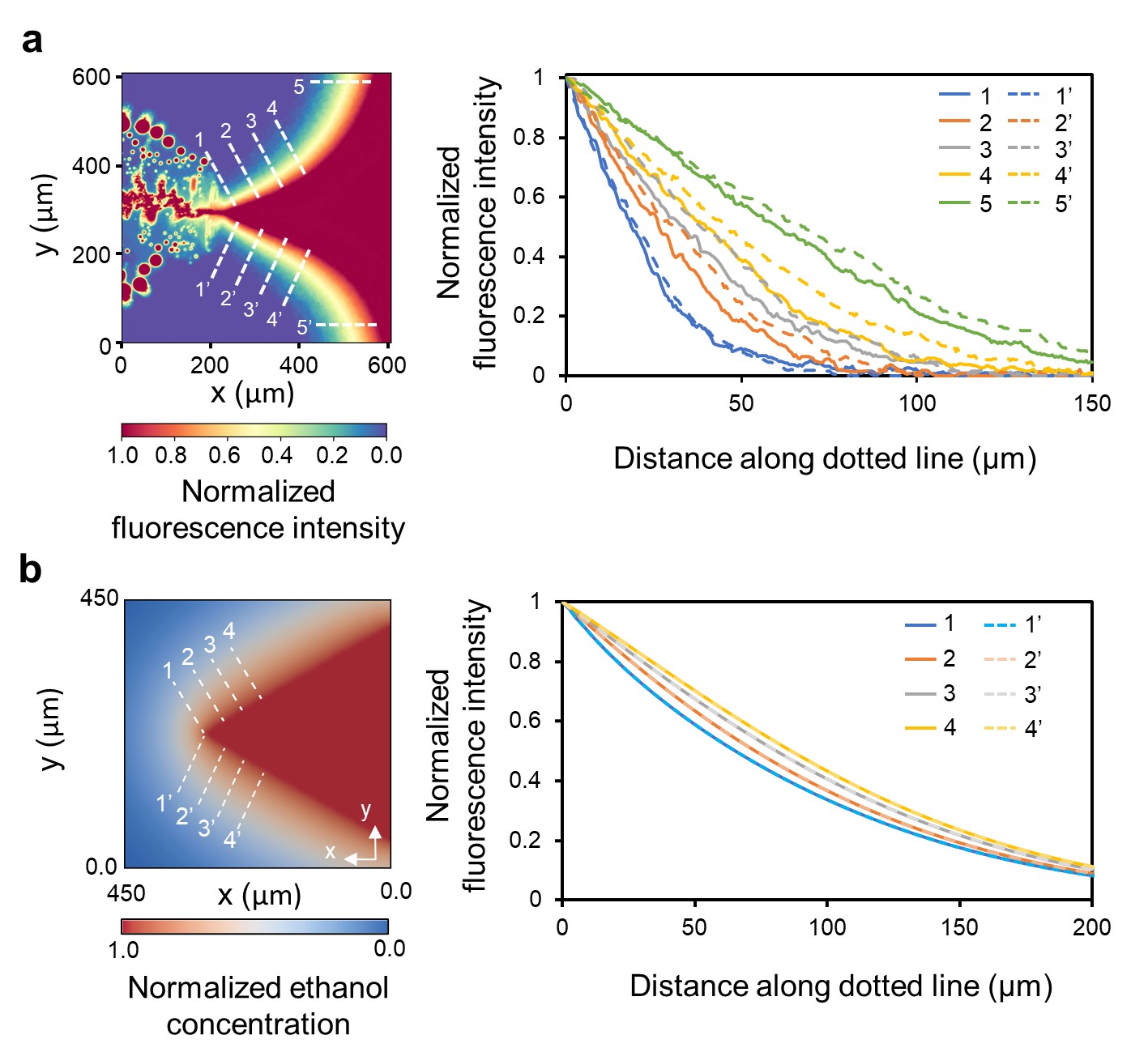}
    \caption{Concentration profiles are symmetric along the y-axis}
    \label{symmetry}
\end{figure}

\newpage
\begin{figure}[ht]
    \centering
    \includegraphics{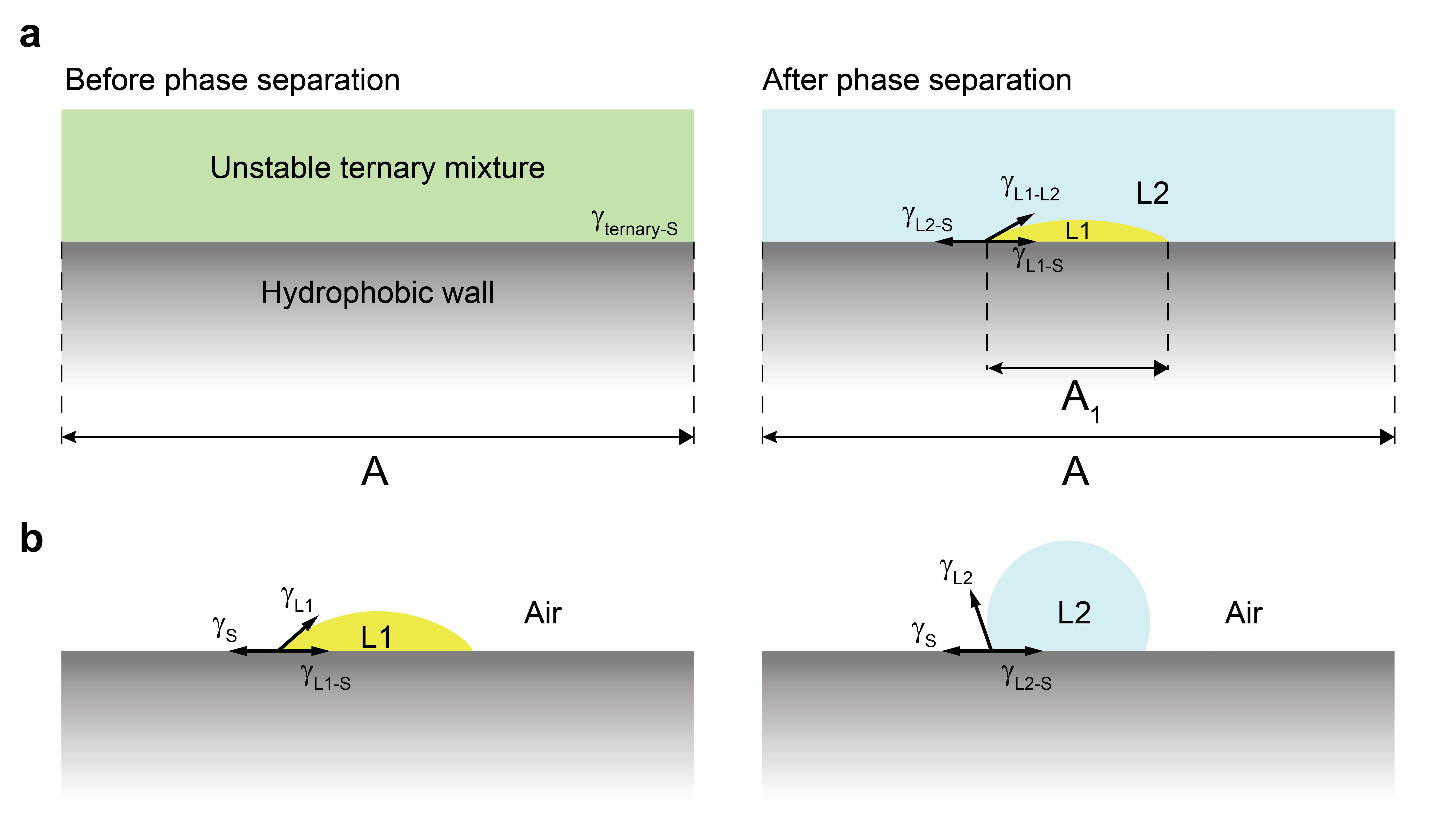}
    \caption{\textbf{a.} Schematic showing the interfaces before and after phase separation and their respective interfacial tensions. \textbf{b.} Schematic showing the force balance on oil and water droplets on the hydrophobic substrate surrounded in air.}
    \label{energy}
\end{figure}

\bibliography{ref_cej}